\documentclass[10pt,journal,twoside]{IEEEtran}
\usepackage{multirow}
\usepackage{amssymb}
\usepackage[dvips]{graphicx}
\usepackage{amsmath}
\usepackage{amsthm}
\usepackage{latexsym,bm}
\usepackage{color}
\usepackage{subfigure}
\usepackage{longtable}
\usepackage{accents}
\usepackage{cite}
\usepackage{enumerate}
\usepackage{arydshln}

\usepackage[ruled]{algorithm}
\usepackage{algorithmic}

\usepackage{diagbox}
\makeatletter
\def\widebar{\accentset{{\cc@style\underline{\mskip10mu}}}}
\def\Widebar{\accentset{{\cc@style\underline{\mskip8mu}}}}
\makeatother

\allowdisplaybreaks
\theoremstyle{plain}

\theoremstyle{definition}
\theoremstyle{definition} 

\begin{document}

\title{{Performance Analysis of a Two-Hop Relaying LoRa System}
\thanks{This work was supported in part by the NSF of China (Nos. 62071129, 62071131, 61771149, U2001203, 61871136, 61871132); in part by the Open Research Fund of the State Key Laboratory of Integrated Services Networks under Grant ISN22-23; in part by the NSF of Guangdong Province under Grant 2019A1515011465; in part by the Science and Technology Program of Guangzhou under Grant 201904010124; in part by the Research Project of the Education Department of Guangdong Province under Grant 2017KTSCX060 and Grant 2017KZDXM028; in part by the Graduate Education and Innovation Project of Guangdong Province under Grant 2020SQXX12; and in part by the Guangdong Innovative Research Team Program under Grant 2014ZT05G157.}
\thanks{W.~Xu, G.~Cai, and Y.~Fang are with the School of Information Engineering, Guangdong University of Technology, China (e-mail: xw\_yang@outlook.com, \{caiguofa2006,fangyi\}@gdut.edu.cn).}
\thanks{G. Chen is with the Department of Electrical Engineering, City University
of Hong Kong, Hong Kong, China (e-mail: eegchen@cityu.edu.hk).}}
\author{Wenyang Xu, Guofa Cai, {\em Member, IEEE}, Yi Fang, {\em Member, IEEE}, and Guanrong Chen, {\em Life Fellow, IEEE}}


\maketitle
\begin{abstract}
The conventional LoRa system is not able to sustain long-range communication over fading channels.
To resolve the challenging issue, this paper investigates a two-hop opportunistic amplify-and-forward relaying LoRa system. Based on the best relay-selection protocol, the analytical and asymptotic bit error rate (BER), achievable diversity order, coverage probability, and throughput of the proposed system are derived over the Nakagami-$m$ fading channel.
Simulative and numerical results show that although the proposed system reduces the throughput compared to the conventional LoRa system, it can significantly improve BER and coverage probability.
Hence, the proposed system can be considered as a promising platform for low-power, long-range and highly reliable wireless-communication applications.

\end{abstract}

\section{Introduction} \label{sect:review}
Emerging from a concept connecting different kinds of objects to the Internet, Internet of Things (IoT) has initiated a revolution worldwide \cite{6740844}.
Some researchers dedicate to enhance the reliability of the communication systems by studying wireless information and power transfer (SWIPT) \cite{9184069}, network coding \cite{8884235} and error-correction techniques \cite{8269289}.
While other researchers are trying to start a new leading era, in order to meet the requirement of wide-area connectivity for low-power and low-data-rate devices, different low-power wide-area network (LPWAN) technologies have been developed to complement traditional cellular and short-range wireless technologies in IoT applications \cite{7815384}. Among them, LoRa wide-area network (LoRaWAN), developed by the LoRa Alliance, is one of the leading wireless standards for IoT devices. LoRa modulation, as the physical-layer technology of LoRaWAN, has inherent advantages for chirp spread spectrum (CSS) modulation, which is robust against interference, multipath fading and Doppler effect\cite{9178997}. Hence, LoRa has drawn considerable attention from both academia and industry.

Recently, several works verified that the performances of LoRa are affected by fading \cite{8392707}, interference \cite{8903531}, and its ALOHA-based protocol \cite{8693695}. Therefore, research efforts have been dedicated to improving the performances of the conventional LoRa network, e.g., increasing the capacity and spectral efficiency by increasing the density of LoRa gateways \cite{8728243}, and to improving its scalability and the reliability by designing effective protocols \cite{9020119}. Still, more effort is needed to enhance the network reliability and extend the network coverage in the industrial scenario.

Since the multi-relay architectures are developed to boost the performance of the system \cite{8019817,8693678}, there are some progresses in applying relaying communication to LoRa networks via advanced protocol design\cite{8048465,8326735,8703036,9136688}.
To construct a robust multi-hop LoRa network, a combination of the LoRa physical-layer standard with concurrent transmission protocol was proposed in \cite{8048465}. In \cite{8326735}, a LoRa mesh system was presented to increase packet delivery ratio without deploying additional gateways.
Moreover, a synchronous LoRa mesh protocol was proposed in \cite{8703036} to improve transmission quality and reduce packet error for underground applications.
By considering the link quality between end nodes and the gateway, a two-hop real-time LoRa protocol was devised in \cite{9136688}, which assigns nodes to different data transmission periods, providing a reliable and energy-efficient scheme.

With the above motivation, this paper proposes a two-hop opportunistic amplify-and-forward (AF) relaying LoRa system. Since the Nakagami-$m$ fading channel represents a wide variety of realistic environment with a fading parameter $m$, the performance of the proposed system is analyzed in terms of bit error rate (BER), diversity order, coverage probability and throughput over a Nakagami-$m$ fading channel. Simulation results are presented to verify the accuracy of the theoretical analysis and show the significant BER and coverage probability improvements at the cost of decreasing the throughput.

\section{System Model} \label{sect:system model}
This section first briefly introduces some basic principles of the LoRa modulation and demodulation, and then presents the two-hop relaying LoRa system model.
\vspace{-0.6cm}
\subsection{LoRa Modulation} \label{subsect:conv_LoRa}

\begin{figure*}[tbp]
\setlength{\abovecaptionskip}{-0.1cm} 
\setlength{\belowcaptionskip}{-1cm} 
\center
\includegraphics[width=7.047in,height=1.277in]{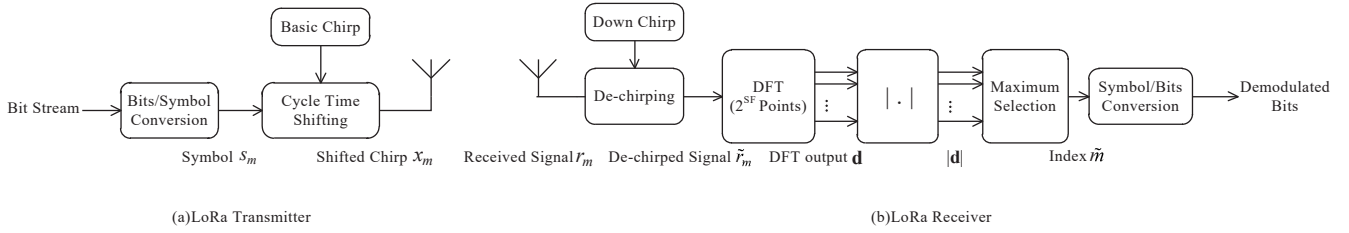}
\caption{A block diagram of the LoRa transceiver.}
\label{fig:LoRatrans}
\vspace{-0.1cm}
\end{figure*}

The transceiver structure of the LoRa system is shown in Fig.~\ref{fig:LoRatrans}.
At the LoRa transmitter, every $SF$ (i.e., the spread factor in LoRa) bits from bit streams are converted to a decimal number $m$, which governs the cyclic time shift of the LoRa chirp signal. Accordingly, a LoRa symbol signal is generated by applying cyclic time shift to the basic chirp signal ${{x}_{0}}(n)$, which is given by
\begin{equation}
\label{eq:lorachirp}
{{x}_{m}}(n)=\sqrt{\frac{1}{{{2}^{SF}}}}\exp \left[ j2\pi \cdot \frac{{{\left( \left( n+m \right)\bmod {{2}^{SF}} \right)}^{2}}}{{{2}^{SF+1}}} \right],
\end{equation}
where $m\in \mathbf{M}=\left\{ 0,1,...,{{2}^{SF}}-1 \right\}$ is a decimal number carried by the LoRa chirp signal. It is noted that the frequency offset of ${{x}_{m}}(n)$ is related to the time shift, i.e., ${{f}_{m}}=B\cdot {m}/{{{2}^{SF}}}$, which sets the starting frequency, where $B$ denotes the bandwidth of LoRa signal. With a slope of ${{{B}^{2}}}/{{{2}^{SF}}}$, the frequency increases from ${{f}_{m}}$ to $B$ and then folds to 0, thereafter it continues increasing to ${{f}_{m}}$ in a symbol interval.

Over a time-invariant and frequency-flat channel, the received LoRa symbol ${{r}_{m}}\left( n \right)$ is written as
\begin{equation}
\label{eq:reclorasig}
{{r}_{m}}\left( n \right)\text{=}h\sqrt{P}{{x}_{m}}\left( n \right)+z\left( n \right),
\end{equation}
where $P$ is the transmitting power, $h$ is the complex envelope amplitude, ${{x}_{m}}\left( n \right)$ is a LoRa symbol with unit energy, and $z\left( n \right)$ denotes the complex additive white Gaussian noise (AWGN).

To demodulate the received LoRa symbol signal ${{r}_{m}}\left( n \right)$, a de-chirped signal ${{\tilde{r}}_{m}}\left( n \right)$ is used to multiply the down chirp with the received signal, given by
\begin{equation}
\label{eq:declorasig}
{{\tilde{r}}_{m}}\left( n \right)={{r}_{m}}\left( n \right)\times x_{0}^{*}\left( n \right),
\end{equation}
where $x_{0}^{*}\left( n \right)$ denotes the complex conjugate of the basis signal ${{x}_{0}}\left( n \right)$.
Then, a ${{2}^{SF}}$-point Discrete Fourier Transform (DFT) is performed for ${{\tilde{r}}_{m}}\left( n \right)$. After applying several algebraic manipulations, the ${{2}^{SF}}$ DFT outputs are given by
\begin{equation}
\label{eq:dftoutput}
\begin{aligned}
   {{\mathbf{d}}_{m}}\!\!\left( k \right)\!\!&=\!\frac{\left| h \right|\!\!\sqrt{P}}{{{2}^{SF}}}\!\exp\! \left(\!\!\frac{{j2\pi {m}^{2}}}{{{2}^{SF+1}}} \!\!+\!\! {j{\varphi _h}}\!\!\right)\!\!\! \sum\limits_{n=0}^{{{2}^{SF}}\!\!-\!1}\!\!\!{\exp\! \left(\!\!\frac{j\pi\!\left( \!m\!-\!k\! \right)\!n\!}{{{2}^{SF-1}}} \!\right)}\!\!+\!\!Z\!\left( k \right),
\end{aligned}
\end{equation}
where $k$ and $m$ are integers, $|.|$ denotes the absolute value, ${\varphi _h}$ is the corresponding phase shift, and $Z\left( k \right)$ is the complex AWGN. The magnitude of ${{\mathbf{d}}_{m}}\left( k \right)$ is given by
\begin{equation}
\label{eq:dftmag}
\left| {{\mathbf{d}}_{m}}\left( k \right) \right|=\left\{ \begin{matrix}
   \left| \left| h \right|\sqrt{P}\exp\! \left(j{\varphi _m}\right)+Z\left( k \right) \right|&,k=m  ~\\
   \left| Z\left( k \right) \right|&,k\ne m,  \\
\end{matrix} \right.
\end{equation}
where ${\varphi _m}$ denotes the phase term as ${\varphi _m}=\frac{{2\pi {k}^{2}}}{{{2}^{SF+1}}}+ {{\varphi _h}}$.
The detected symbol $\tilde{m}$ can be estimated by selecting the index of the maximum DFT output, given by
\begin{equation}
\label{eq:dftmaxindex}
\tilde{m}=\underset{k\in \mathbf{M}}{\mathop{\arg \max }}\,\left( \left| {{\mathbf{d}}_{m}}\left( k \right) \right| \right).
\end{equation}

\subsection{Two-Hop AF Relaying LoRa System}\label{subsect:protocolchannel}

\begin{figure}[tbp]
\center
\vspace{-0.5cm}
\includegraphics[width=3.583in,height=1.933in]{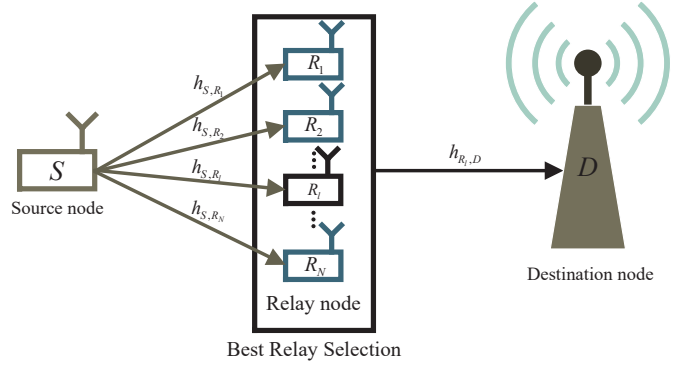}
\vspace{-0.4cm}
\caption{A two-hop AF relaying LoRa system.}
\setlength{\abovecaptionskip}{-0.1cm} 
\setlength{\belowcaptionskip}{-1cm} 
\label{fig:LoRaRelaySystem}
\vspace{-0.2cm}
\end{figure}

A two-hop opportunistic AF relaying LoRa system is considered, as shown in Fig.~\ref{fig:LoRaRelaySystem}, which includes one source node $S$, $N$ candidate relay nodes (i.e., ${{R}_{1}},{{R}_{2}},...,{{R}_{N}}$), and one destination node $D$ (LoRa gateway). The source node $S$ broadcasts a signal to all the relays over the source-to-relay ($S$-$R$) links. Then,  the selected best relay amplifies the received signal, and forwards it to the destination node. Note that, there is no direct link between source node and destination node because of poor channel condition.

It should be noted that, the proposed system performs selection strategy using the protocol described in \cite{9136688}, which can select the best relay by examine the link quality (i.e., SNR) when constructing the parent-child tree between nodes and gateway during network initialization.

In phase \uppercase\expandafter{\romannumeral1}, a LoRa signal is broadcast by the source node and then received by $N$ relays over a fading channel. The received signal at the ${l}$-th relay is given by
\begin{equation}
\label{eq:relayphase1}
y_{S,R_{l} } =\sqrt{P_{S} } h_{S,R_{l} } x+z_{S,R_{l} } ,{\rm \; }l=1,2,...,N,
\end{equation}
where $x$ is the transmitted LoRa signal from the source, ${{P}_{S}}$ is the transmit power of the source, ${{h}_{S,{{R}_{l}}}}$ denotes the fading coefficient of the channel from source $S$ to relay ${{R}_{l}}$, and ${{z}_{S,{{R}_{l}}}}$ is the corresponding noise at the ${l}$-th relay, modeled as ${{z}_{S,{{R}_{l}}}}\sim \mathcal{C}\mathcal{N}\left( 0,{{N}_{S,{{R}_{l}}}} \right)$.

In phase \uppercase\expandafter{\romannumeral2}, the best relay is selected by the maximum signal-to-noise ratio (SNR) strategy, which is analyzed in Section~\ref{sect:perfomnanceanalysis}. Specifically, the ${l}$-th relay amplifies the signal received and forwards it to the destination node. Thus, the received signal at destination is given by
\begin{equation}
\label{eq:relayphase2}
{{y}_{D}}={{h}_{{{R}_{l}},D}}A{{y}_{S,{{R}_{l}}}}+{{z}_{{{R}_{l}},D}},
\end{equation}
where $A$ is the amplification factor, given by $A=\sqrt{\frac{{{P}_{{{R}_{l}}}}}{\left| {{h}_{S,{{R}_{l}}}} \right|^{2}{{P}_{S}}+{{N}_{S,{{R}_{l}}}}}}$, ${{P}_{{{R}_{l}}}}$ is the transmit power of the relay, ${{h}_{{{R}_{l}},D}}$ denotes the fading coefficient of the ${{R}_{l}}$-$D$ channel, and the noise at the destination is modeled as ${{z}_{{{R}_{l}},D}}\sim \mathcal{C}\mathcal{N}\left( 0,{{N}_{{{R}_{l}},D}} \right)$.

In this paper, all the fading coefficients are assumed to be independent and identically distributed (i.i.d.) Nakagami-$m$ random variables (RVs).
Without loss of generality, assume that all the noise are i.i.d. complex Gaussian distribution RVs with zero mean and variance ${{N}_{0}/{2^{SF}}}$, i.e., ${{N}_{S,{{R}_{l}}}}={{N}_{{{R}_{l}},D}}={{N}_{0}/{2^{SF}}}$. Thus, the instantaneous SNR of $S$-${{R}_{l}}$ and ${{R}_{l}}$-$D$ links are given by ${{\gamma }_{S,{{R}_{l}}}}\!=\!\frac{{{P}_{S}}{{\left| {{h}_{S,{{R}_{l}}}} \right|}^{2}}}{{{N}_{0}/{2^{SF}}}}$ and 	 ${{\gamma }_{{{R}_{l}},D}}\!=\!\frac{{{P}_{R}}{{\left| {{h}_{{{R}_{l}},D}} \right|}^{2}}}{{{N}_{0}/{2^{SF}}}}$, respectively.
When a path-loss model is considered in the relaying LoRa system \cite{9244125,8728243}, the corresponding average SNR are ${{\bar{\gamma }}_{S,{{R}_{l}}}}\!=\!\frac{{{P}_{S}}d_{S,{{R}_{l}}}^{-\alpha }}{{{N}_{0}/{2^{SF}}}}$ and ${\bar{\gamma }_{{{R}_{l}},D}}=\frac{{{P}_{R}}d_{{{R}_{l}},D}^{-\alpha }}{{{N}_{0}/{2^{SF}}}}$, respectively, where $d$ is the internode distance and $\alpha$ is the path loss exponent ranging from 2 to 6.

\section{Performance Analysis} \label{sect:perfomnanceanalysis}

This section presents the analytical and asymptotic BER, achievable diversity order and throughput of the proposed system over a Nakagami-$m$ fading channel.

Due to the Nakagami-$m$ fading, for any link between $S$-${{R}_{l}}$ or ${{R}_{l}}$-$D$, the probability density function (PDF) of the instantaneous SNR ${{\gamma }_{j,k}}$, is given by\cite{simon2005digital}
\begin{equation}
\label{eq:p2pnakapdf}
{{f}_{{{\gamma }_{j,k}}}}(r)={{\left( \frac{{{m}_{j,k}}}{{{{\bar{\gamma }}}_{j,k}}} \right)}^{{{m}_{j,k}}}}\frac{1}{\Gamma \left( {{m}_{j,k}} \right)}{{r}^{{{m}_{j,k}}-1}}{{e}^{-\frac{{{m}_{j,k}}}{{{{\bar{\gamma }}}_{j,k}}}r}},
\end{equation}
where ${{m}_{j,k}}$ denotes the Nakagami-$m$ fading parameter of link $j$-$k$, where $j\in \{S,{{R}_{l}}\}, k\in \{{{R}_{l}},D\}$ and $\Gamma \left( \cdot  \right)$ is the gamma function.
The exact instantaneous end-to-end SNR ${{\gamma }_{{{t}_{l}}}}$ of a single path, through the ${l}$-th relay (i.e., $S$-${{R}_{l}}$-$D$), is given by\cite{1244790}
\begin{equation}
\label{eq:singlepathsnr}
{{\gamma }_{{{t}_{l}}}}=\frac{{{\gamma }_{S,{{R}_{l}}}}{{\gamma }_{{{R}_{l}},D}}}{{{\gamma }_{S,{{R}_{l}}}}+{{\gamma }_{{{R}_{l}},D}}+1}.
\end{equation}
\vspace{-0.2cm}
\subsection{Analytical BER Analysis } \label{subsect:AnalyticalBER}

According to Section \ref{subsect:protocolchannel}, the best relay node is selected to maximize the instantaneous end-to-end SNR ${{\gamma }_{{{t}_{\max }}}}$ at the destination. Hence, one has
\begin{equation}
\label{eq:maxsnr}
{{\gamma }_{{{t}_{\max }}}}=\underset{l=1,2,...,N}{\mathop{\max }}\,\left( {{\gamma }_{{{t}_{l}}}} \right).
\end{equation}

According to the order statistics\cite{hogg2005introduction}, the PDF of the end-to-end SNR ${{\gamma }_{{{t}_{\max }}}}$ is given by
\begin{equation}
\label{eq:maxpdf}
{{f}_{{{\gamma }_{{{t}_{\max }}}}}}(r)=\sum\limits_{k=1}^{N}{{{f}_{{{\gamma }_{{{t}_{k}}}}}}}(r)\prod\limits_{\begin{smallmatrix}
 l=1 \\
 l\ne k
\end{smallmatrix}}^{N}{{{F}_{{{\gamma }_{{{t}_{l}}}}}}}(r),
\end{equation}
and the CDF of the end-to-end SNR ${{\gamma }_{{{t}_{\max }}}}$ is given by
\begin{equation}
\label{eq:maxcdf}
{{F}_{{{\gamma }_{{{t}_{\max }}}}}}(r)=\prod\limits_{k=1}^{N}{{{F}_{{{\gamma }_{{{t}_{k}}}}}}}(r).
\end{equation}

The exact instantaneous end-to-end PDF and CDF of ${{\gamma }_{{{t}_{l}}}}$ over a Nakagami-$m$ fading channel is derived in \cite{6220222}, which can be substituted into Eq.~(\ref{eq:maxpdf}) and Eq.~(\ref{eq:maxcdf}) to obtain the exact closed-form expressions for the PDF and the CDF of the end-to-end SNR ${{\gamma }_{{{t}_{\max }}}}$.

According to \cite{simon2005digital}, an error is occurred in LoRa symbol detection when
\begin{equation}
\underset{k\in \mathbf{M},k\ne m}{\mathop{\max }}\,\left( \left| {{\mathbf{d}}_{m}}\left( k \right) \right| \right)>\left|{{\mathbf{d}}_{m}}\left( k\left| k=m \right. \right)\right|,
\end{equation}
where $\underset{k\in \mathbf{M},k\ne m}{\mathop{\max }}\,\left( \left| {{\mathbf{d}}_{m}}\!\left( k \right) \right| \right)$ can be approximated as a constant $\sqrt{{{N}_{0}}\!\cdot\! {{H}_{{{2}^{SF}}\!-1}}}$ \cite{8392707}, because it exhibits a very low coefficient of variation, where $H_{x}$ denotes the $x$-th harmonic number.
Hence, the conditional error probability for LoRa symbol detection $P_{e|{{h }}}$ corresponding to $h$ is approximated as
\begin{equation}
\label{eq:errorprobforlora}
  {{P}_{e|{{h }}}}\approx\Pr \left[ \sqrt{{{N}_{0}}\!\cdot\! {{H}_{{{2}^{SF}}\!-1}}}>\left|{{\mathbf{d}}_{m}}\left( k\left| k=m \right. \right) \right| \right],
\end{equation}
where ${{\mathbf{d}}_{m}}\left( k\left| k=m \right. \right) = \sqrt{P}\left| h \right|\exp\! \left(j{\varphi _m}\right)+Z\left( k \right)$, which follows a complex Gaussian distribution as
\begin{equation}
\begin{aligned}
{{\mathbf{d}}_{m}}&\left( k\left| k  = m \right. \right)\\
\!\!\!&\sim \mathcal{C}\mathcal{N}\!\!\left ( \sqrt{P} \left | h \right |  cos\left (\varphi _m \right )+ j\sqrt{P} \left | h \right | sin\left ( \varphi _m \right ), {{{N}_{0}}} \right ).
\end{aligned}
\end{equation}

According to the basic properties of the complex Gaussian distribution, $\left|{{\mathbf{d}}_{m}}\left( k\left| k  = m \right. \right)\right|$ follows a Rician distribution with the shape parameter ${{\kappa }_{\mathbf{d}}}=\frac{P{{\left| h \right|}^{2}}}{{{N}_{0}}}$, which can be further  approximated as a Gaussian random variable with the mean ${\sqrt{P}{{\left| h \right|}}}$ and the variance ${{{N}_{0}}}/{2}$ \cite{8392707}. Hence, Eq. (\ref{eq:errorprobforlora}) is calculated as
\begin{equation}
\label{eq:errorprobforlora-Qfunc}
{{P}_{e|{{h }}}}\approx Q\left( \frac{\sqrt{P}\left| h \right|-\sqrt{{{N}_{0}}{{H}_{{{2}^{SF}}-1}}}}{\sqrt{{{N}_{0}}/2}} \right),
\end{equation}
where $Q\left( x \right)=\frac{1}{\sqrt{2\pi }}\int_{x}^{\infty }{\exp \left( -\frac{{{u}^{2}}}{2} \right)}du$ is the Gaussian $Q$ function.
According to \cite{Proakis2007}, the conditional BER ${{P}_{b|\gamma }}$ corresponding to $\gamma$ can be calculated as
\begin{equation}
\label{eq:conditionBER}
{{P}_{b|{{\gamma }}}}\approx 0.5\times {Q}\left( \sqrt{2\gamma }-\sqrt{2{{H}_{{{2}^{SF}}-1}}} \right),
\end{equation}
where $\gamma = \frac{P{{\left| h \right|}^{2}}}{{{N}_{0}}}$ denotes the SNR.

Accordingly, the analytical BER ${{P}_{b}}$ is expressed as a single-fold integral, i.e.,
\begin{equation}
\label{eq:analyticalBER}
{{P}_{b}}\approx 0.5\times \int\limits_{0}^{\infty }{{Q}\left( \sqrt{2{{\gamma }_{{{t}_{\max }}}}\left( r \right)}-\sqrt{2{{H}_{{{2}^{SF}}-1}}} \right){{f}_{{{\gamma }_{{{t}_{\max }}}}}}(r)dr}.
\end{equation}

To give more insight, we further derive the closed-form BER expression by the asymptotic BER analysis in Section~\ref{subsect:AsymptoticBER}.


\vspace{-0.3cm}
\subsection{Asymptotic BER Analysis } \label{subsect:AsymptoticBER}

Now, the asymptotic BER performance of the proposed system is analyzed. The single path exact instantaneous end-to-end SNR ${{\gamma }_{{{t}_{l}}}}$ can be estimated by its upper bound, given by \cite{4155634}
\begin{equation}
\gamma _{{{t}_{l}}}^{up}=\min \left( {{\gamma }_{S,{{R}_{l}}}},{{\gamma }_{{{R}_{l}},D}} \right).
\end{equation}

Hence, the PDF of $\gamma _{{{t}_{l}}}^{up}$ is expressed as
\begin{eqnarray} \nonumber
\label{eq:boundpdf}
   {{f}_{\gamma _{{{t}_{l}}}^{up}}}(r)\!\!\!&&=\!\!\left[ {{\!\left( \! \frac{{{m}_{S,{{R}_{l}}}}}{{{{\bar{\gamma }}}_{S,{{R}_{l}}}}} \!\!\right)}^{\!\!{{m}_{S,{{R}_{l}}}}}}\!\!\!\!\!\!\!{{r }^{{{m}_{S,{{R}_{l}}}}-1}}{{e}^{-\frac{{{m}_{S,{{R}_{l}}}}r }{{{{\bar{\gamma }}}_{S,{{R}_{l}}}}}}}\Gamma\! \left( \! {{m}_{{{R}_{l}},D}},\frac{{{m}_{{{R}_{l}},D}}r }{{{{\bar{\gamma }}}_{{{R}_{l}},D}}}  \!\!\right) \right. \\ \nonumber
  &&\left.+{{\left(\!\! \frac{{{m}_{{{R}_{l}},D}}}{{{{\bar{\gamma }}}_{{{R}_{l}},D}}} \!\!\right)}^{\!\!{{m}_{{{R}_{l}},D}}}}\!\!\!\!\!\!\!\!\!{{r }^{{{m}_{{{R}_{l}}\!,\!D}}-1}}{{e}^{\!-\frac{{{m}_{{{R}_{l}},D}}r }{{{{\bar{\gamma }}}_{{{R}_{l}},D}}}}}\!\Gamma\!\! \left( \! {{m}_{S,{{R}_{l}}}},\frac{{{m}_{S,{{R}_{l}}}}r }{{{{\bar{\gamma }}}_{S,{{R}_{l}}}}} \!\right) \!\right] \\
  &&\times \left[ \Gamma ({{m}_{S,{{R}_{l}}}})\Gamma ({{m}_{{{R}_{l}},D}}) \right]^{-1},
\end{eqnarray}
where $\Gamma \left( \cdot ,\cdot  \right)$ is the upper incomplete gamma function\cite[Eq.~(8.350.2)]{jeffrey2007table}, and $\Gamma \left( \cdot  \right)$ denotes the gamma function\cite[ Eq.~(8.310.1)]{jeffrey2007table}. Using a Maclaurin series expansion, the PDF of $\gamma _{{{t}_{l}}}^{up}$ is approximated as\cite{1221802}
\begin{equation}
\label{eq:approxpdf}
{{f}_{\gamma _{{{t}_{l}}}^{up}}}(r)\approx {{a}_{l}}{{r}^{{{t}_{l}}}},r\to 0,
\end{equation}
where ${{t}_{l}}=\min \left( {{m}_{S,{{R}_{l}}}},{{m}_{{{R}_{l}},D}} \right)-1$ and ${{a}_{l}}$ is the first non-zero Maclaurin coefficient of ${{f}_{\gamma _{{{t}_{l}}}^{up}}}(r)$ when $r$ tends to 0. Hence, the factor of the ${{t}_{l}}$-th-order derivative of ${{f}_{\gamma _{{{t}_{l}}}^{up}}}(r)$ is expressed as
\begin{equation}
\label{eq:pdfmacseries}
{{a}_{l}}=\frac{1}{{{t}_{l}}!}\frac{{{\partial }^{{{t}_{l}}}}}{\partial {{r}^{{{t}_{l}}}}}{{f}_{{{\gamma }_{{{t}_{l}}}}}}(0).
\end{equation}

Using the chain rule for differentiating composite functions, Eq. (\ref{eq:pdfmacseries}) can be obtained as
\begin{eqnarray}
\label{eq:pdfseries}
{{a}_{l}} \!\!& = &\!\! \left\{\!\! \begin{array}{*{35}{l}}
   \frac{{{\left( \frac{{{m}_{S,{{R}_{l}}}}}{{{{\bar{\gamma }}}_{S,{{R}_{l}}}}} \right)}^{{{m}_{S,{{R}_{l}}}}}}}{\Gamma \left( {{m}_{S,{{R}_{l}}}} \right)}, & {{m}_{S,{{R}_{l}}}}\!<\!{{m}_{{{R}_{l}},D}}  \\
   \frac{{{\left( \frac{{{m}_{S,{{R}_{l}}}}}{{{{\bar{\gamma }}}_{S,{{R}_{l}}}}} \right)}^{\!\!{{m}_{S,{{R}_{l}}}}}}}{\Gamma \left( {{m}_{S,{{R}_{i}}}} \right)}\!+\!\frac{{{\left( \frac{{{m}_{{{R}_{l}},D}}}{{{{\bar{\gamma }}}_{{{R}_{l}},D}}} \right)}^{\!\!{{m}_{{{R}_{l}},D}}}}}{\Gamma \left( {{m}_{{{R}_{l}},D}} \right)},\!\! & {{m}_{S,{{R}_{l}}}} \!=\! {{m}_{{{R}_{l}},D}}  \\
   \frac{{{\left( \frac{{{m}_{{{R}_{l}},D}}}{{{{\bar{\gamma }}}_{{{R}_{l}},D}}} \right)}^{{{m}_{{{R}_{l}},D}}}}}{\Gamma \left( {{m}_{{{R}_{l}},D}} \right)}, & {{m}_{S,{{R}_{l}}}}\!>\!{{m}_{{{R}_{l}},D}}.  \\
\end{array} \right.
\end{eqnarray}

It should be noted that the ${t}$-th $\left( t<{{t}_{l}} \right)$-order derivative of ${{f}_{\gamma _{{{t}_{l}}}^{up}}}(0)$ is null. The CDF of
$\gamma _{{{t}_{l}}}^{up}$ can be derived by integrating ${{f}_{\gamma _{{{t}_{l}}}^{up}}}(r)$ over $r$ as
\begin{equation}
\label{eq:cdfseries}
{{F}_{\gamma _{{{t}_{l}}}^{up}}}\left( r \right)\approx \frac{{{a}_{l}}{{r}^{{{t}_{l}}+1}}}{\min \left( {{m}_{S,{{R}_{l}}}},{{m}_{{R}_{l},D}} \right)}.
\end{equation}

Using Eq.~(\ref{eq:maxpdf}), the upper bound of PDF of the end-to-end SNR $\gamma _{{{t}_{\max }}}^{up}$ is given by
\begin{equation}
\label{eq:maxsnrboundpdf}
{{f}_{\gamma _{{{t}_{\max }}}^{up}}}(r)=\sum\limits_{k=1}^{N}{{{a}_{k}}{{r}^{{{t}_{k}}}}}\prod\limits_{\begin{smallmatrix}
 l=1 \\
 l\ne k
\end{smallmatrix}}^{N}{\frac{{{a}_{l}}{{r}^{{{t}_{l}}+1}}}{\min \left( {{m}_{S,{{R}_{l}}}},{{m}_{{R}_{l},D}} \right)}},
\end{equation}
where ${{t}_{l}}=\min \left( {{m}_{S,{{R}_{l}}}},{{m}_{{R}_{l},D}} \right)-1$.

According to the conditional BER ${{P}_{b|{{\gamma }}}}$ in Section~\ref{subsect:AnalyticalBER}, one can obtain the asymptotic BER $P_{b}^{asy}$, given by
\begin{equation}
\label{eq:asymberintegral}
P_{b}^{asy}\approx0.5\times \int_{0}^{\infty }{{Q}\left( \sqrt{2r}-V \right){{f}_{\gamma _{{{t}_{\max }}}^{up}}}(r)dr},
\end{equation}
where $V=\sqrt{2{{H}_{{{2}^{SF}}-1}}}$. Substituting Eq.~(\ref{eq:maxsnrboundpdf}) into Eq.~(\ref{eq:asymberintegral}) and performing some algebraic manipulations, one has

\begin{eqnarray}
\label{eq:fullexpanasymber}
  P_{b}^{asy} \!=\!\! &0.5&\!\!\times \sum\limits_{k = 1}^{N}{\frac{\prod\limits_{l = 1}^{N}{{{a}_{l}}}}{\prod\limits_{\begin{smallmatrix}
 l = 1 \\ \nonumber
 l\ne k
\end{smallmatrix}}^{N}{\min \left( {{m}_{S,{{R}_{l}}}},{{m}_{{R}_{l},{{D}}}} \right)}}} \\
  &\times&\!\!\!\!\!\!\int_{0}^{\infty }\!\!\!\!{{Q}\!\left(\! \sqrt{2r}\!-\!V \!\right)\!{{r}^{\left( \sum\limits_{k  =  1}^{N}{\min \left( {{m}_{S,{{R}_{l}}}},{{m}_{{R}_{l},{D}}} \right)} \!\right)-1}}}\!\!dr.
\end{eqnarray}

Using the upper bound of the Gaussian $Q$ function in \cite{Proakis2007}, i.e., ${Q}\left( x \right)\le \frac{1}{2}{{e}^{-\frac{{{x}^{2}}}{2}}}$, one can further derive an approximation as
\begin{equation}
\label{eq:asymberQfuncapprox}
P_{b}^{asy}\approx \frac{1}{4}\times \sum\limits_{k=1}^{N}{\frac{\prod\limits_{l=1}^{N}{{{a}_{l}}}}{\prod\limits_{\begin{smallmatrix}
 l=1 \\
 l\ne k
\end{smallmatrix}}^{N}{\min \left( {{m}_{S,{{R}_{l}}}},{{m}_{{{R}_{l}},D}} \right)}}}\!\times\! \!\int_{0}^{\infty }\!\!{{\!\!{e}^{-\frac{{{\left( \sqrt{2r}-V \right)}^{2}}}{2}}}{{r}^{\Delta }}dr},
\end{equation}
where $\Delta \!=\!\left( \sum\limits_{k=1}^{N}{\min \left( {{m}_{S,{{R}_{l}}}},{{m}_{{R}_{l},{D}}} \right)} \right)\!-\!1$.

Applying a substitution with $x=\sqrt{2r}-V$ and using the binomial theorem, the integral in Eq.~(\ref{eq:asymberQfuncapprox}) is simplified as
\begin{equation}
\label{eq:integralexpand}
\begin{aligned}
  & \int_{0}^{\infty }{{{e}^{-\frac{{{\left( \sqrt{2r}-V \right)}^{2}}}{2}}}{{r}^{\Delta }}dr} \\
 & ={{2}^{-\Delta }}\times \sum\limits_{j=0}^{2\Delta +1}{\left( \begin{matrix}
   2\Delta +1  \\
   j  \\
\end{matrix} \right){{V}^{2\Delta +1 -j}}}\times \int\limits_{-V}^{\infty }{{{e}^{-\frac{{{x}^{2}}}{2}}}{{x}^{j}}dx}. \\
\end{aligned}
\end{equation}

Applying \cite[Eq.~(3.381.8), Eq.~(3.381.9)]{jeffrey2007table} and performing some manipulations, the integral in Eq.~(\ref{eq:integralexpand}) is computed as
\begin{equation}
\label{eq:asymberintegralapprox}
\int\limits_{-V}^{\infty }{{{e}^{-\frac{{{x}^{2}}}{2}}}{{x}^{j}}dx}={{\left( -1 \right)}^{j}}\times {{\left( \frac{1}{2} \right)}^{v-1}}\!\!\!\times \left( \gamma \left( v,{{{V}^{2}}}/{2}\; \right)+\Gamma \left( v \right) \right),
\end{equation}
where $v=\frac{j+1}{2}$, $\gamma (\alpha ,x)=\int_{0}^{x}{{{e}^{-t}}}{{t}^{\alpha -1}}dt$, and $\Gamma (z)=\int_{0}^{\infty }{{{t}^{z-1}}}{{e}^{-t}}dt$.

Finally, the closed-form expression for the asymptotic BER $P_{b}^{up}$ is obtained by using Eq.~(\ref{eq:asymberQfuncapprox}), Eq.~(\ref{eq:integralexpand}), and Eq.~(\ref{eq:asymberintegralapprox}), as
\begin{equation}
\label{eq:asymberclosedform}
\begin{aligned}
   P_{b}^{asy}&\approx \frac{1}{4}\times \frac{\prod\limits_{l=1}^{N}{{{a}_{l}}}}{\sum\limits_{k=1}^{N}{\prod\limits_{\begin{smallmatrix}
 l=1 \\
 l\ne k
\end{smallmatrix}}^{N}{\min \left( {{m}_{S,{{R}_{l}}}},{{m}_{{{R}_{l}},{{D}}}} \right)}}}\times {{2}^{-\Delta }} \\
 & \times \sum\limits_{j=0}^{2\Delta +1}{\left( \begin{matrix}
   2\Delta +1  \\
   j  \\
\end{matrix} \right)}\left[ \begin{aligned}
  & {{V}^{2\Delta +1 -j}}\times {{\left( -1 \right)}^{j}}\times {{\left( {}^{1}\!\!\diagup\!\!{}_{2}\; \right)}^{v-1}} \\
 & \times \left( \gamma \left( v,{{{V}^{2}}}/{2}\; \right)+\Gamma \left( v \right) \right) \\
\end{aligned} \right].
\end{aligned}
\end{equation}

The asymptotic BER expression in Eq.~(\ref{eq:asymberclosedform}) can be used to analyze the achievable diversity order of the proposed system. Following the definition of diversity order in \cite{1197843}, using Eq.~(\ref{eq:asymberclosedform}) and letting ${{\bar{\gamma }}_{S,{{R}_{l}}}}={{\bar{\gamma }}_{{{R}_{l}},D}}=\bar{\gamma }$, one can calculate the diversity order of the proposed system as
\begin{eqnarray}
\label{eq:diversity} \nonumber
  {{G}_{d}}& = & \underset{\text{SNR}\to \infty }{\mathop{\lim }}\,-\frac{\log P_{b}^{asy}(\text{SNR})}{\log \text{SNR}} \\
 & = & \sum\limits_{l = 1}^{N}{\min \left( {{m}_{S,{{R}_{l}}}},{{m}_{{R}_{l},{D}}} \right)}.
\end{eqnarray}

\vspace{-0.6cm}
\subsection{Coverage Probability Analysis} \label{subsect:coverageprobana}

The coverage probability of the LoRa relaying network can be defined as the probability of the event that the end-to-end SNR is greater that the a target SNR $\psi$, which is given by
\begin{equation}
\label{eq:coverageprobdefine}
{P}_{cov}= \text{Pr}\left( {{\gamma }_{{{t}_{\max }}}} > \psi \right).
\end{equation}
It indicates that the coverage probability is the complementary CDF of the SNR ${{\gamma }_{{{t}_{\max }}}}$.

Using Eq.~(\ref{eq:maxcdf}), the coverage probability is expressed as
\begin{equation}
\label{eq:coverageprobdefine}
{P}_{cov}= 1-{{F}_{{{\gamma }_{{{t}_{\max }}}}}}(\psi).
\end{equation}

\vspace{-0.6cm}
\subsection{Throughput Analysis} \label{subsect:BERcomparison}

In a wireless system, a significant performance criterion is the throughput, which can be defined as the number of the correct bits received per unit time\cite{5934344}. Thus, the throughput for a LoRa system is defined as
\begin{equation}
\label{eq:throughput}
{{R}_{t}}=\left[ L\cdot SF\cdot \left( 1-PER \right) \right]/{{T}_{t}},
\end{equation}
where $L$ is the number of symbols in each packet, $PER$ is the packet error rate, and ${{T}_{t}}$ denotes the transmission period for a given system $t\in \left\{ \text{Conv}\text{., Relay} \right\}$.

For the LoRa modulation, $PER$ can be expressed as
\begin{equation}
\label{eq:packeterrorrate}
{PER}=1-{{\left( 1-{{P}_{e}} \right)}^{L}},
\end{equation}
where ${{P}_{e}}=2{{P}_{b}}$ is the symbol error rate.

Since the proposed system performs a best relay selection scheme, only one relay is selected to retransmit the information to the destination, i.e., two transmission periods are needed regardless of the number of relays.
The transmission periods for the proposed system and the conventional LoRa system are ${{T}_{\text{Relay}}}=2L \cdot {{T}_{sym}\left(SF \right)}$ and ${{T}_{\text{Conv}\text{.}}}=L \cdot {{T}_{sym}\left(SF \right)}$, respectively, where ${{T}_{sym}\left(SF \right)}=2^{SF}{T_{sam}}$ denotes the LoRa symbol interval, which is a function of the spread factor $SF$, and ${T_{sam}}$ denotes the sample interval. Hence, their corresponding throughput can be calculated as ${{R}_{\text{Conv}\text{.}}}\!=\!SF\left[ {{\left( 1-{{P}_{e}} \right)}^{L}} \right]/{{{T}_{sym}\left(SF \right)}}$
and ${{R}_{\text{Relay}}}\!=\!SF\left[ {{\left( 1-{{P}_{e}} \right)}^{L}} \right]/{2{{T}_{sym}\left(SF \right)}}$, respectively.

\section{Results and Discussions} \label{sect:resultanddiscussion}	

\begin{figure}[tbp]
\center
\includegraphics[width=2.34in,height=1.747in]{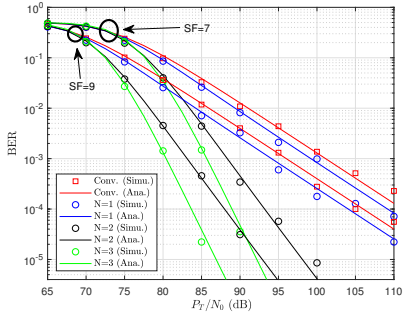}
\vspace{-0.4cm}
\caption{Analytical and simulated BER curves of the proposed system with different numbers of relays over a Rayleigh fading channel. The spread factor is set to $SF=7$ and $9$, respectively.}
\label{fig:BERvsPtn0}
\vspace{-0.6cm}
\end{figure}

In this section, BER, coverage probability, and throughput of the proposed system over Nakagami-$m$ channel are evaluated. Unless otherwise specified, in simulations, the total power in the system is set to 14 dBm \cite{8693695}, the distance between the source node and the destination node is set to 2 kilometers (km). To be fair, the total powers of the proposed system and the conventional system are set to the same,  and we plot the BER versus $P_{T}/N_{0}$ for better comparison. The average channel gains are related to the path loss $d_{j,k}^{-\alpha }$, where the path loss exponent $\alpha$ is set to 2.65 with the practical city environment considered in the simulation \cite{8292708}.

Fig.~\ref{fig:BERvsPtn0} compares the analytical and simulated BER results of the conventional LoRa system (without relay) and the proposed system over a Rayleigh fading channel.
As the link average SNR is set to ${{\bar{\gamma }}_{S,{{R}_{l}}}}={{\bar{\gamma }}_{{{R}_{l}},D}}$, the asymptotic results are very tight in the medium-and-high-SNR regime. ~\footnote{It should be noted that, the existing works for BER analysis of the LoRa system do not consider a realistic pass loss model \cite{8392707,8903531}, which lead to a relatively low SNR regime.}
It can be observed that, as $N$ increases, although the BER performance of the proposed system can be improved, the gap between the systems with $N$ and $N-1$ ($N>1$) gradually decreases at the same BER.
Moreover, as $SF$ gets larger, the BER performance is getting better.
Furthermore, the limiting slopes of BER are affected by the relay number, which confirms the analysis of  the diversity order in Eq.~(\ref{eq:diversity}).

\begin{figure}[tbp]
\center
\includegraphics[width=2.34in,height=1.747in]{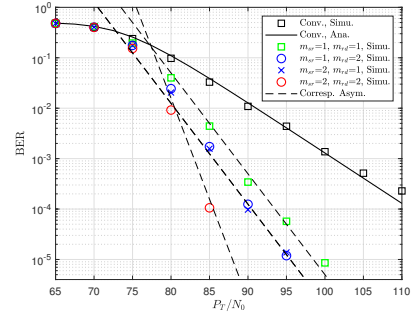}
\vspace{-0.4cm}
\caption{Asymptotic and simulated BER curves of the proposed system over different channel profiles. The proposed system is considered in the cases of $N=2$ and $SF=7$.}
\label{fig:BERcompdiffpara}
\vspace{-0.4cm}
\end{figure}

Fig.~\ref{fig:BERcompdiffpara} compares the asymptotic and simulated BER results of the proposed system over different fading channels, where the effect of the channel condition (i.e., the fading parameter of the Nakagami-$m$ fading channel) is illustrated under the assumption of  ${{\bar{\gamma }}_{S,{{R}_{l}}}}={{\bar{\gamma }}_{{{R}_{l}},D}}$.
Both symmetric and asymmetric Nakagami-$m$ fading links are considered for the case of relay number $N=2$.
It can be observed that, the asymptotic BER results are very tight in the medium-and-high-SNR
regime.
Moreover, the BER performance of the proposed system is improved as fading mitigates.
Furthermore, the dash lines and the blue markers, verify that the diversity order is determined by the more severely faded link (i.e., $\min \left( {{m}_{S,{{R}_{l}}}},{{m}_{{R}_{l},{{D}}}} \right)$ from Eq.~(\ref{eq:diversity})).

\begin{figure}[tbp]
\center
\includegraphics[width=2.325in,height=1.7903in]{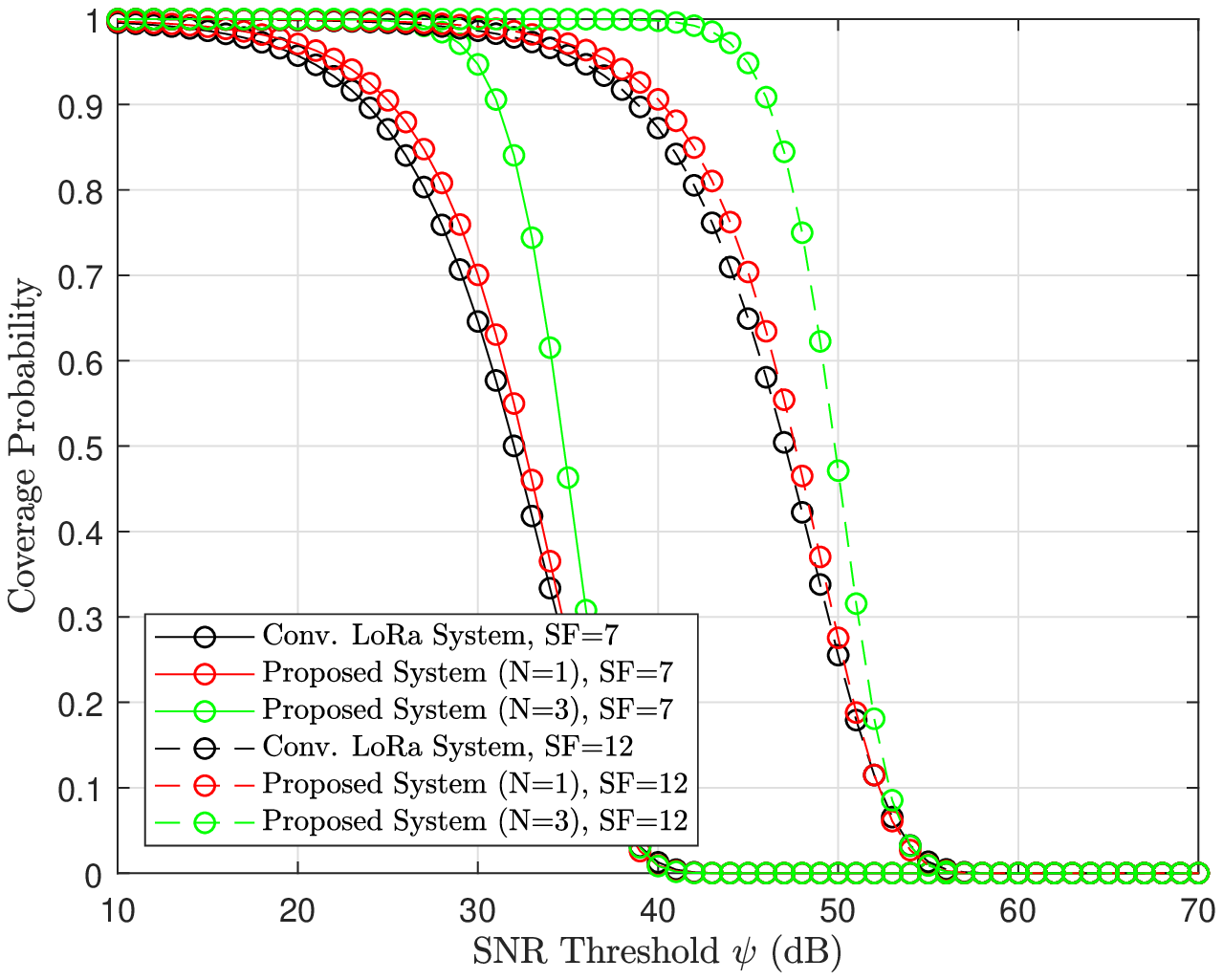}
\vspace{-0.45cm}
\caption{Coverage Probability for different LoRa systems over a Nakagami-$m$ fading channel. The proposed system is considered in the cases of $N=1$ and $3$, respectively. The spread factor is set to $SF=7\text{ and } 12$, respectively.}
\label{fig:coverageprob}
\vspace{-0.5cm}
\end{figure}

Fig.~\ref{fig:coverageprob} illustrates the coverage probability for the conventional LoRa system and the proposed systems over the Nakagami-$m$ ($m=1$) fading channel, where $P_{T}/N_{0}$ is set to $100$ dB.
It can be observed that the proposed system has better coverage probability than the conventional LoRa system under small-and-medium SNR threshold.
Moreover, increasing the number of relays can significantly boost the coverage probability.
For example, at the SNR threshold of $30$~dB, the coverage probabilities for the proposed system are about $1.08$ and $1.38$ times of the conventional LoRa system when $SF=7$, in the scenarios with $N=1,3$, respectively.
Furthermore, when $SF$ gets larger, the coverage probability is getting higher.


\begin{figure}[tbp]
\center
\includegraphics[width=2.2896in,height=1.8536in]{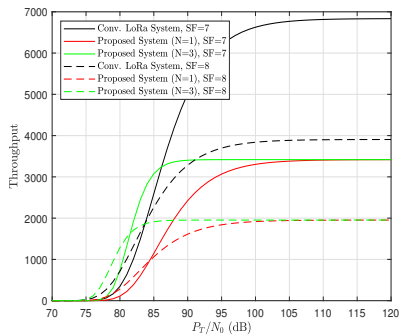}
\vspace{-0.1cm}
\caption{Throughput of different LoRa systems over a Nakagami-$m$ fading channel. The proposed system is considered in the cases of $N=1$ and $3$, respectively. The spread factor is set to $SF=7$ and $12$, respectively.}
\label{fig:throughput}
\vspace{-0.5cm}
\end{figure}

Fig.~\ref{fig:throughput} shows the throughput of the conventional LoRa system and the proposed system over the Nakagami-$m$ ($m=1$) fading channel. It can be observed that the throughput of the proposed system is better than the conventional LoRa system in the low-and-medium-SNR regime in multi-relay scenario.
Moreover, as the number of relays increases, the throughput of the proposed system can reach its limitation faster.
Meanwhile, when $SF$ gets larger, the throughput of the proposed system is getting lower.
Furthermore, referring to Figs. \ref{fig:BERvsPtn0}, \ref{fig:coverageprob}, and \ref{fig:throughput}, and compared to the conventional LoRa system, the proposed system improves the BER performance and coverage probability. However, a lower throughput is gained in the high-SNR regime.
\vspace{-0.2cm}

\section{Conclusion} \label{sect:conclusion}
In this paper, a two-hop AF relaying LoRa system has been investigated over Nakagami-$m$ fading channels  as a preliminary work. The analytical and asymptotic BER expressions, achievable diversity order, coverage probability, and throughput of the proposed system have been analyzed.
In addition, compared with the conventional LoRa system, the coverage probability of the proposed system can be remarkably improved at the cost of decreasing the throughput.
Thanks to the aforementioned advantages, the proposed system appears to be a promising framework for low-power, long-range and highly reliable wireless-communication applications. In future work, we will consider the self-interference of LoRa signal in multi-relay scenario to further analyze the system performance.
\vspace{-0.2cm}

\bibliographystyle{IEEEtran}
\bibliography{IEEEabrv,Relaying-LoRa-System_Conf_Init}

\begin{thebibliography}{10}
\providecommand{\url}[1]{#1}
\csname url@samestyle\endcsname
\providecommand{\newblock}{\relax}
\providecommand{\bibinfo}[2]{#2}
\providecommand{\BIBentrySTDinterwordspacing}{\spaceskip=0pt\relax}
\providecommand{\BIBentryALTinterwordstretchfactor}{4}
\providecommand{\BIBentryALTinterwordspacing}{\spaceskip=\fontdimen2\font plus
\BIBentryALTinterwordstretchfactor\fontdimen3\font minus
  \fontdimen4\font\relax}
\providecommand{\BIBforeignlanguage}[2]{{%
\expandafter\ifx\csname l@#1\endcsname\relax
\typeout{** WARNING: IEEEtran.bst: No hyphenation pattern has been}%
\typeout{** loaded for the language `#1'. Using the pattern for}%
\typeout{** the default language instead.}%
\else
\language=\csname l@#1\endcsname
\fi
#2}}
\providecommand{\BIBdecl}{\relax}
\BIBdecl

\bibitem{6740844}
A.~{Zanella}, N.~{Bui}, A.~{Castellani}, L.~{Vangelista}, and M.~{Zorzi},
  ``Internet of things for smart cities,'' \emph{IEEE Internet Things J.},
  vol.~1, no.~1, pp. 22--32, Feb. 2014.

\bibitem{9184069}
G.~Cai, Y.~Fang, P.~Chen, G.~Han, G.~Cai, and Y.~Song, ``Design of an
  {MISO}-{SWIPT}-aided code-index modulated multi-carrier m-{DCSK} system for
  e-health {IoT},'' \emph{IEEE J. Sel. Areas Commun.}, vol.~39, no.~2, pp.
  311--324, Feb. 2021.

\bibitem{8884235}
P.~Chen, Z.~Xie, Y.~Fang, Z.~Chen, S.~Mumtaz, and J.~J. P.~C. Rodrigues,
  ``Physical-layer network coding: An efficient technique for wireless
  communications,'' \emph{IEEE Netw.}, vol.~34, no.~2, pp. 270--276, Mar. 2020.

\bibitem{8269289}
Y.~Fang, G.~Han, G.~Cai, F.~C.~M. Lau, P.~Chen, and Y.~L. Guan, ``Design
  guidelines of low-density parity-check codes for magnetic recording
  systems,'' \emph{IEEE Commun. Surveys Tuts.}, vol.~20, no.~2, pp. 1574--1606,
  Secondquarter 2018.

\bibitem{7815384}
U.~{Raza}, P.~{Kulkarni}, and M.~{Sooriyabandara}, ``Low power wide area
  networks: An overview,'' \emph{IEEE Commun. Surveys Tuts.}, vol.~19, no.~2,
  pp. 855--873, Secondquarter 2017.

\bibitem{9178997}
X.~{Xia}, Y.~{Zheng}, and T.~{Gu}, ``{FTrack}: Parallel decoding for {LoRa}
  transmissions,'' \emph{IEEE/ACM Trans. Netw.}, vol.~28, no.~6, pp.
  2573--2586, Dec. 2020.

\bibitem{8392707}
T.~{Elshabrawy} and J.~{Robert}, ``Closed-form approximation of {LoRa}
  modulation {BER} performance,'' \emph{IEEE Commun. Lett.}, vol.~22, no.~9,
  pp. 1778--1781, Sep. 2018.

\bibitem{8903531}
O.~{Afisiadis}, M.~{Cotting}, A.~{Burg}, and A.~{Balatsoukas-Stimming}, ``On
  the error rate of the {LoRa} modulation with interference,'' \emph{IEEE
  Trans. Wireless Commun.}, vol.~19, no.~2, pp. 1292--1304, Feb. 2020.

\bibitem{8693695}
J.~{Ortín}, M.~{Cesana}, and A.~{Redondi}, ``Augmenting {LoRaWAN} performance
  with listen before talk,'' \emph{IEEE Trans. Wireless Commun.}, vol.~18,
  no.~6, pp. 3113--3128, Jun. 2019.

\bibitem{8728243}
Z.~{Qin}, Y.~{Liu}, G.~Y. {Li}, and J.~A. {McCann}, ``Performance analysis of
  clustered {LoRa} networks,'' \emph{IEEE Trans. Veh. Technol.}, vol.~68,
  no.~8, pp. 7616--7629, Aug. 2019.

\bibitem{9020119}
Q.~L. {Hoang}, W.~{Jung}, T.~{Yoon}, D.~{Yoo}, and H.~{Oh}, ``A real-time
  {LoRa} protocol for industrial monitoring and control systems,'' \emph{IEEE
  Access}, vol.~8, pp. 44\,727--44\,738, Mar. 2020.

\bibitem{8019817}
Y.~Fang, S.~C. Liew, and T.~Wang, ``Design of distributed protograph ldpc codes
  for multi-relay coded-cooperative networks,'' \emph{IEEE Trans. Wireless
  Commun.}, vol.~16, no.~11, pp. 7235--7251, Nov. 2017.

\bibitem{8693678}
Y.~Fang, P.~Chen, G.~Cai, F.~C. Lau, S.~C. Liew, and G.~Han,
  ``Outage-limit-approaching channel coding for future wireless communications:
  Root-protograph low-density parity-check codes,'' \emph{IEEE Veh. Technol.
  Mag.}, vol.~14, no.~2, pp. 85--93, Jun. 2019.

\bibitem{8048465}
C.~{Liao}, G.~{Zhu}, D.~{Kuwabara}, M.~{Suzuki}, and H.~{Morikawa}, ``Multi-hop
  {LoRa} networks enabled by concurrent transmission,'' \emph{IEEE Access},
  vol.~5, pp. 21\,430--21\,446, Sep. 2017.

\bibitem{8326735}
H.~{Lee} and K.~{Ke}, ``Monitoring of large-area {IoT} sensors using a {LoRa}
  wireless mesh network system: Design and evaluation,'' \emph{IEEE Trans.
  Instrum. Meas.}, vol.~67, no.~9, pp. 2177--2187, Sep. 2018.

\bibitem{8703036}
C.~{Ebi}, F.~{Schaltegger}, A.~{Rüst}, and F.~{Blumensaat}, ``Synchronous
  {LoRa} mesh network to monitor processes in underground infrastructure,''
  \emph{IEEE Access}, vol.~7, pp. 57\,663--57\,677, Apr. 2019.

\bibitem{9136688}
H.~P. {Tran}, W.~{Jung}, T.~{Yoon}, D.~{Yoo}, and H.~{Oh}, ``A two-hop
  real-time {LoRa} protocol for industrial monitoring and control systems,''
  \emph{IEEE Access}, vol.~8, pp. 126\,239--126\,252, Jul. 2020.

\bibitem{9244125}
Q.~M. Qadir, ``Analysis of the reliability of lora,'' \emph{IEEE Commun.
  Lett.}, vol.~25, no.~3, pp. 1037--1040, March 2021.

\bibitem{simon2005digital}
M.~K. Simon and M.-S. Alouini, \emph{Digital communication over fading
  channels}.\hskip 1em plus 0.5em minus 0.4em\relax John Wiley \& Sons, 2005.

\bibitem{1244790}
M.~Hasna and M.-S. Alouini, ``End-to-end performance of transmission systems
  with relays over rayleigh-fading channels,'' \emph{IEEE Trans. Wireless
  Commun.}, vol.~2, no.~6, pp. 1126--1131, Nov. 2003.

\bibitem{hogg2005introduction}
R.~V. Hogg, J.~McKean, and A.~T. Craig, \emph{Introduction to mathematical
  statistics}.\hskip 1em plus 0.5em minus 0.4em\relax Pearson Education, 2005.

\bibitem{6220222}
S.~S. {Soliman} and N.~C. {Beaulieu}, ``Exact analysis of dual-hop {AF} maximum
  end-to-end {SNR} relay selection,'' \emph{IEEE Trans. Commun.}, vol.~60,
  no.~8, pp. 2135--2145, Aug. 2012.

\bibitem{Proakis2007}
J.~G. Proakis and M.~Salehi, \emph{Digital Communications}, 5th~ed.\hskip 1em
  plus 0.5em minus 0.4em\relax McGraw Hill, 2007.

\bibitem{4155634}
S.~{Ikki} and M.~H. {Ahmed}, ``Performance analysis of cooperative diversity
  wireless networks over nakagami-m fading channel,'' \emph{IEEE Commun.
  Lett.}, vol.~11, no.~4, pp. 334--336, Apr. 2007.

\bibitem{jeffrey2007table}
A.~Jeffrey and D.~Zwillinger, \emph{Table of integrals, series, and
  products}.\hskip 1em plus 0.5em minus 0.4em\relax Elsevier, 2007.

\bibitem{1221802}
Z.~Wang and G.~B. Giannakis, ``A simple and general parameterization
  quantifying performance in fading channels,'' \emph{IEEE Trans. Commun.},
  vol.~51, no.~8, pp. 1389--1398, Aug. 2003.

\bibitem{1197843}
L.~Zheng and D.~N.~C. Tse, ``Diversity and multiplexing: a fundamental tradeoff
  in multiple-antenna channels,'' \emph{IEEE Trans. Inf. Theory}, vol.~49,
  no.~5, pp. 1073--1096, May 2003.

\bibitem{5934344}
Z.~Wang, Q.~Peng, and L.~B. Milstein, ``Multi-user resource allocation for
  downlink multi-cluster multicarrier {DS} {CDMA} system,'' \emph{IEEE Trans.
  Wireless Commun.}, vol.~10, no.~8, pp. 2534--2542, Aug. 2011.

\bibitem{8292708}
P.~Jorke, S.~Bocker, F.~Liedmann, and C.~Wietfeld, ``Urban channel models for
  smart city {IoT}-networks based on empirical measurements of {LoRa}-links at
  433 and 868 {MHz},'' in \emph{Proc. IEEE 28th Annu. Int. Symp. Pers. Indoor
  Mobile Radio Commun. (PIMRC)}.\hskip 1em plus 0.5em minus 0.4em\relax {IEEE},
  Oct. 2017, pp. 1--6.

\end{thebibliography}

\end{document}